\documentclass[prl,aps,graphicx,superscriptaddress,reprint]{revtex4-1}
\usepackage{graphicx}
\draft 

\begin{document}


\title{Universal Substrate Effect on the Superconductivity of FeSe Monolayer Films} 



\author{S. N. Rebec}
\altaffiliation{These two authors contributed equally to this work.}
\affiliation{Department of Applied Physics, Stanford University, Stanford, California 94305, USA}
\affiliation{Stanford Institute for Materials and Energy Sciences, SLAC National Accelerator Laboratory, Menlo Park, California 94025, USA}

\author{T. Jia}
\altaffiliation{These two authors contributed equally to this work.}
\affiliation{Stanford Institute for Materials and Energy Sciences, SLAC National Accelerator Laboratory, Menlo Park, California 94025, USA}
\affiliation{Department of Physics, Stanford University, Stanford, California 94305, USA}

\author{C. Zhang}
\affiliation{Stanford Institute for Materials and Energy Sciences, SLAC National Accelerator Laboratory, Menlo Park, California 94025, USA}
\affiliation{Department of Physics, Stanford University, Stanford, California 94305, USA}

\author{M. Hashimoto}

\affiliation{Stanford Synchrotron Radiation Lightsource, SLAC National Accelerator Laboratory, Menlo Park, California 94025, USA}

\author{D.-H. Lu}

\affiliation{Stanford Synchrotron Radiation Lightsource, SLAC National Accelerator Laboratory, Menlo Park, California 94025, USA}

\author{R. G. Moore}
\affiliation{Stanford Institute for Materials and Energy Sciences, SLAC National Accelerator Laboratory, Menlo Park, California 94025, USA}

\author{Z.-X. Shen}

\affiliation{Department of Applied Physics, Stanford University, Stanford, California 94305, USA}\affiliation{Stanford Institute for Materials and Energy Sciences, SLAC National Accelerator Laboratory, Menlo Park, California 94025, USA}

\affiliation{Department of Physics, Stanford University, Stanford, California 94305, USA}

\date{\today}

\begin{abstract}

To elucidate the mechanisms behind the enhanced $T_c$ in monolayer (1ML) FeSe on SrTiO$_3$ (STO), we grew highly strained 1ML FeSe on the rectangular (100) face of rutile TiO$_2$, and observed the coexistence of replica bands and superconductivity with a $T_c$ of 63 K. From the similar $T_c$ between this system and 1ML FeSe on STO (001), we conclude that strain and dielectric constant are likely unimportant to the enhanced $T_c$ in these systems. A systematic comparison of 1ML FeSe on TiO$_2$ with other systems in the FeSe family shows that while charge transfer alone can enhance $T_c$, it is only with the addition of interfacial electron-phonon coupling that $T_c$ can be increased to the level seen in 1ML FeSe on STO.
\end{abstract}


\maketitle 

Bulk FeSe superconducts at 8 K \cite{Hsu08pnas}. Surprisingly, $T_c$ dramatically increases to $\sim$70K in 1ML FeSe/STO(001) \cite{WangXue12cpl}. Charge transfer from STO is widely believed to play an important role in the superconducting mechanism \cite{HeZhou13nmatt}, which is supported by doping experiments of bulk or multilayer FeSe \cite{MiyataTakahashi15nmat,YeYan15arxiv,WenFeng16ncomm,TangXue15arxiv, LeiChen15arxiv,Shiogai15nphy}. However, the $T_c$ in the doped FeSe systems is less than 50 K, which suggests that there are other substrate related effects contributing to the higher $T_c$ in  1ML FeSe/STO(001). Multiple candidates have been suggested to explain this $T_c$ difference, including interfacial electron-phonon (e-ph) coupling \cite{Lee14nature}, a very high dielectric constant \cite{WangXue12cpl}, and lattice strain \cite{TanFeng13nmat}. Strong, small-q e-ph coupling between carriers in FeSe and optical phonons in STO leads to replica bands observed in angle-resolved photoemission spectroscopy (ARPES) \cite{Lee14nature} measurements. This coupling has been proposed to enhance $T_c$ \cite{RademakerJohnston15NJP}. STO has an extremely high dielectric constant at low temperature\cite{STOepsilon}, which has been suggested to enhance the $T_c$ \cite{WangXue12cpl}. There is also a mismatch between the lattice constant of STO (3.90 \AA) and bulk FeSe (3.76 \AA) that induces a strain on the 1ML FeSe film. Strain could influence superconductivity as $T_c$ is a pressure dependent quantity in bulk FeSe \cite{TanFeng13nmat,Mizuguchiaplet}. 

\begin{figure*}
\centering
\includegraphics[width=0.9\textwidth]{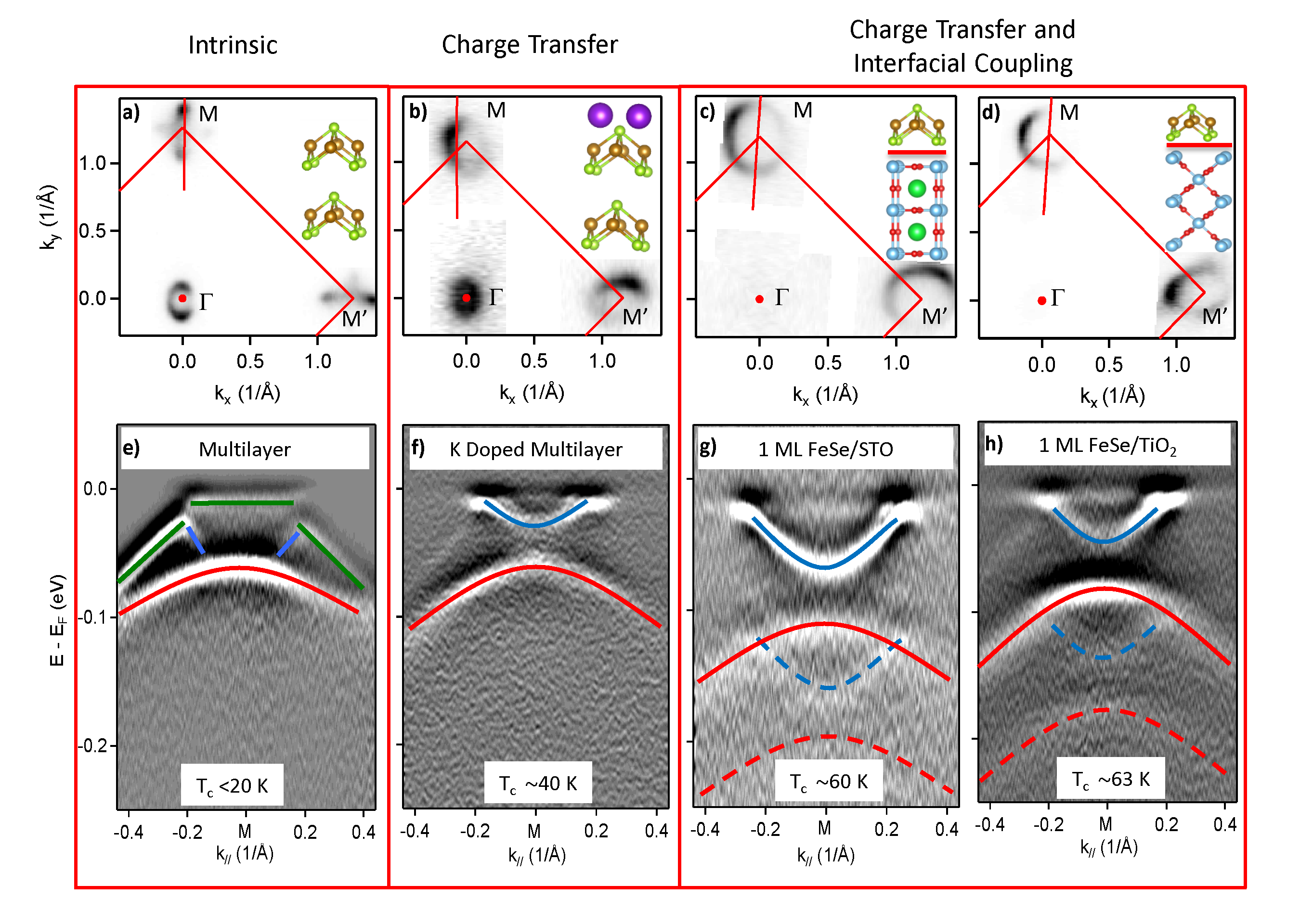}

\caption{\label{fig:1}(a-d): Fermi surface of 60ML FeSe/STO$^{(001)}$, K doped 3ML FeSe/STO$^{(001)}$, 1ML FeSe/STO$^{(001)}$ and 1ML FeSe/TiO$_2$. For (a-c), the spectra at $M^\prime$ are obtained by rotating the data by 90 degrees. Inset: the schematic structure of the respective systems. (e-h): the band structures of respective systems along the cut at M shown in the Fermi surface map above. The band structures shown in (e-h) are second derivatives of the original band with respect to energy.}
\end{figure*}

An ideal substrate to help determine which of the many proposed properties of STO (001) enhance $T_c$ in the 1ML FeSe system is rutile TiO$_2$ (100). Rutile TiO$_2$ is composed of corner-sharing octahedra of oxygen enclosed titanium atoms, while anatase has edge-sharing octahedra. Rutile TiO$_2$ is different from STO in two key ways: dielectric constant and lattice spacing. The low temperature dielectric constant of rutile TiO$_2$ is $\leq$ 260 \cite{Parker61pr_epsilonTiO2}, which is much smaller than STO's dielectric constant of $\sim$10,000 \cite{STOepsilon}. The (100) face of rutile TiO$_2$ is rectangular with lattice spacing a = 2.95 \AA ~and b = 4.59 \AA, while STO (001) is square. Therefore FeSe is expected to have a large, anisotropic strain when grown on rutile TiO$_2$ (100). Despite these differences from STO, previous works show that TiO$_2$ has similar oxygen optical phonon modes with energies of around 100 meV\cite{MCP_TiO2}.

In this work, we systematically compare several previously studied FeSe-based materials, 60ML FeSe/STO$^{(001)}$, K-doped 3ML FeSe/STO$^{(001)}$, 1ML FeSe/STO$^{(001)}$ with a new system, 1ML FeSe on rutile TiO$_2$ (100) (1ML FeSe/TiO$_2^{(100)}$), to determine which substrate effects are central to superconductivity. All the films were grown via molecular beam epitaxy (MBE) and measured with \textit{in-situ} ARPES. Our results indicate that electron-phonon coupling is necessary and sufficient for enhancing the $T_c$ in ML FeSe on STO, beyond charge transfer.

\begin{figure*}
\includegraphics[width=0.9\textwidth]{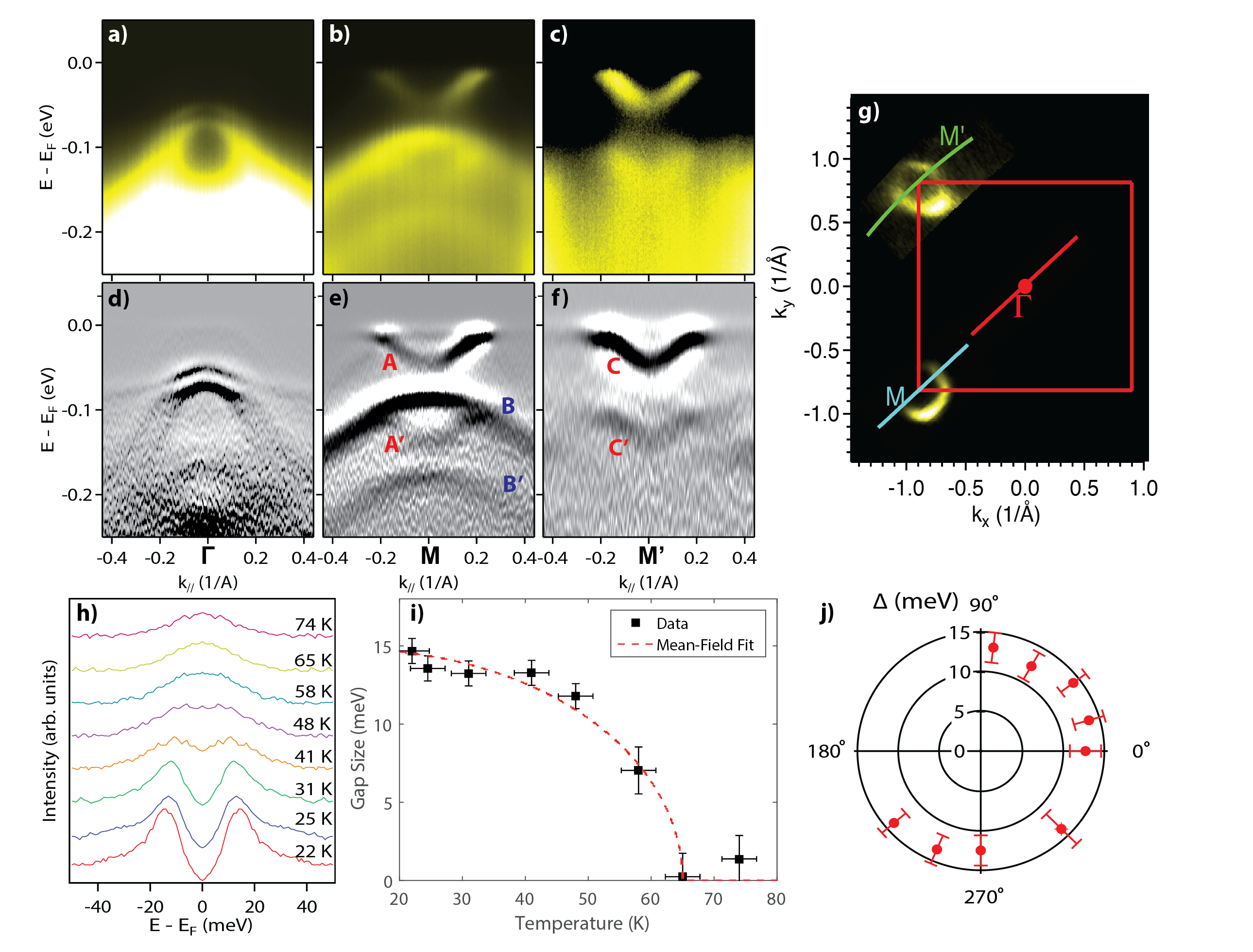}
\caption{\label{fig:2} (a-c): Band structure along high symmetry cuts $\Gamma$, M, and M$^\prime$, and their respective second derivative plots (d-f). (g): Fermi surface map with indication of Brilluion Zone and cut locations. (h) Temperature dependent symmetrized EDC of $k_f$ crossing. (i) Temperature dependent fitted gap size, with mean field fit. (j) Angle dependent gap size around M } 
\end{figure*}

The 1ML FeSe and the K-doped FeSe systems share similar band structures, which are distinct from that of multilayer FeSe, as are shown in the Fermi surface maps (Fig. 1(a) to (d)) and the Brillouin zone-corner cut (Fig. 1(e) to (h)). Multilayer FeSe has hole pockets around $\Gamma$ point and $\lq$dumbbell' shaped spectral weight near M point (Fig. 1(a)). The band structure of multilayer FeSe films is similar from 2 ML to above 60 ML, and also resembles that of bulk FeSe\cite{NakayamaTakahashi15prb}. In comparison, the other three systems have much simpler Fermi surfaces, with no pockets at $\Gamma$ and electron pockets around the M point. In the these three systems, hole bands are observed $\sim$45 meV below the electron bands in the high-symmetry cuts at the M point. The quantitative variation of band positions can be trivially ascribed to the varying doping levels.

Despite the similarities in the band structures of the K-doped and the 1ML films, strong interfacial mode coupling as evidenced by the replica bands, are only observed in the 1ML films grown on STO or TiO$_2$ (Fig. 1(g) and 1(h)). Replica bands are direct copies of primary bands that are rigidly shifted in energy due to small-q, e-ph scattering \cite{Lee14nature}. The energy shift in the replica band is similar in both systems: approximately 100 meV in 1ML FeSe/STO$^{(001)}$ and 90 meV in 1ML FeSe/TiO$_2^{(100)}$.

More detailed measurements of the electronic structure of 1ML FeSe/TiO$_2^{(100)}$ are shown in Fig. 2. From Fig. 2(g) we find that the Brillouin zone is rectangular instead of square, showing a broken $C_4$ symmetry. The lattice constants are measured by Fermi surface map and found to be a= 3.53 \AA ~and b = 3.95 \AA. The Fermi surface consists of elliptical electron pockets at zone corners M and M$^\prime$. Either anisotropic strain or a band-selective matrix element could explain the difference between the shape of electron pockets at M and M$^\prime$.  Overall, the Fermi surface topology, doping level and band structure of 1ML FeSe/TiO$_2^{(100)}$ are similar to those of 1ML FeSe/STO$^{(001)}$.
Replica bands are clearly resolved in the ARPES spectra in high-symmetry cuts at M$^\prime$ and M and are denoted by a prime in (Fig. 2(e)-(f)). They all have an energy shift of $\sim$90 meV from the original band, which is close to the bulk rutile $A_{1g}$ and $B_{2g}$ oxygen optical phonon mode energies\cite{MCP_TiO2}.

The symmetrized energy distribution curve (EDC) (Fig. 2h) from the cut crossing the M point (Fig. 2g) reveals a  superconducting gap of 14 meV at 22 K. The temperature dependent gap was fit using mean field theory \cite{Normanpr_Pheno} and shows a $T_c$ of 63 $\pm$3 K (Fig. 2i). Despite the anisotropy in the stretched Brillouin zone, the gap around M is isotropic within experimental error (Fig. 2j).

1ML FeSe/TiO$_2^{(100)}$ and 1ML FeSe/STO$^{(001)}$ have very similar electronic structure, superconducting gap size, and $T_c$. The similarities between these two systems, suggests that dielectric constant, strain and C4 symmetry likely do not play important roles in the enhancement of superconductivity of the 1ML FeSe on the STO and related substrates. These results are supported by other recent works of 1ML FeSe on other substrates \cite{PengFeng14ncomm, ZhangDing15arxiv, ZhouXue15arxiv, DingXue16arxiv}. 

\begin{figure}
\includegraphics[width=0.5\textwidth]{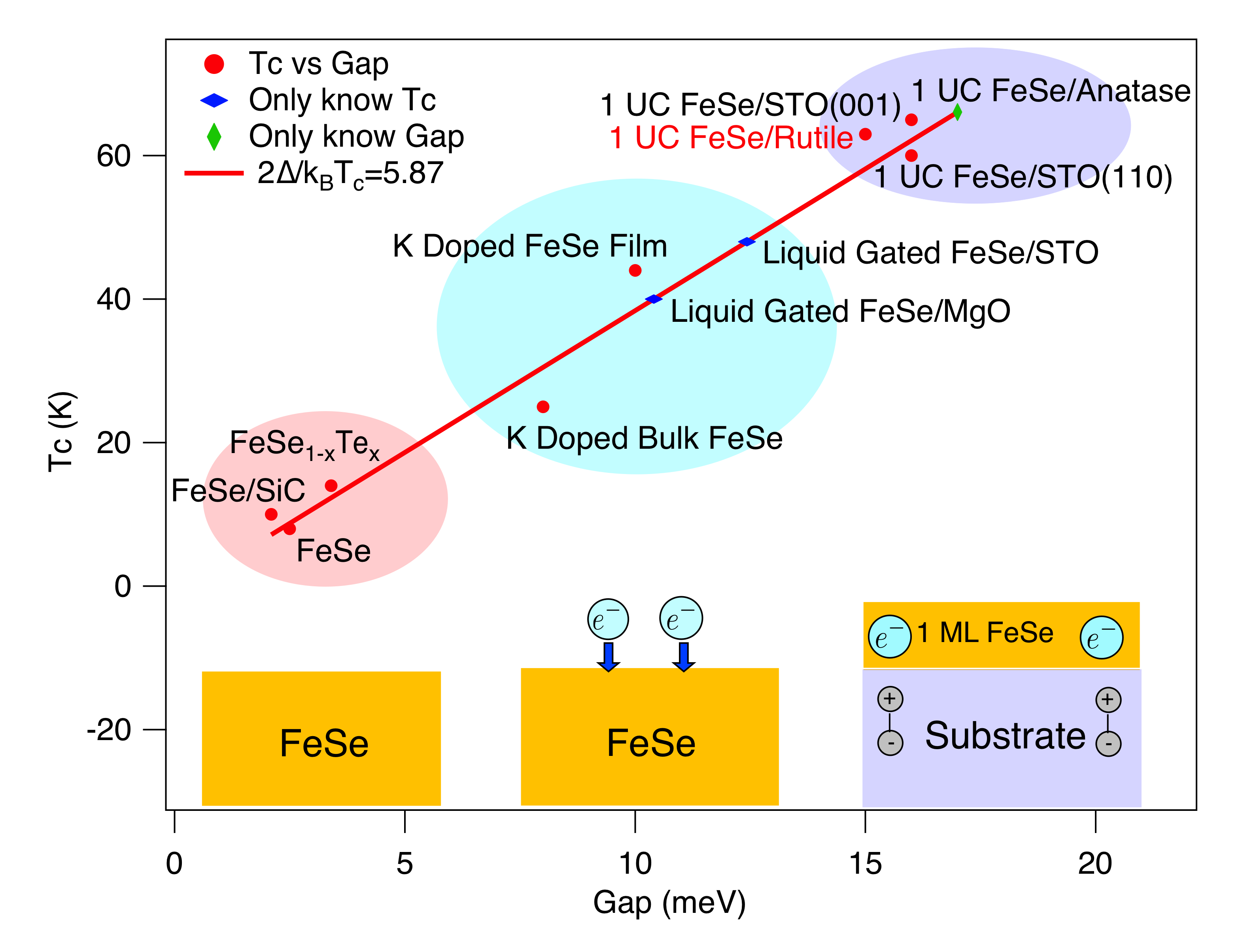}
\caption{\label{fig:3} Plotted are the $T_c$ and maximum superconducting gap of bulk FeSe\cite{Hsu08pnas}, FeTe$_{1-x}$Se$_x$\cite{Noji10jps}, FeSe on SiC\cite{SongXue12prb}, K doped bulk FeSe\cite{YeYan15arxiv}, K doped FeSe film on STO$^{(001)}$\cite{WenFeng16ncomm,TangXue15arxiv}, liquid gated FeSe/MgO\cite{Shiogai15nphy} and FeSe/STO$^{(001)}$\cite{LeiChen15arxiv}, 1ML FeSe/rutile (this work), 1ML FeSe/anatase \cite{DingXue16arxiv}, 1ML FeSe/STO$^{(001)}$\cite{TanFeng13nmat} and 1ML FeSe/STO$^{(110)}$\cite{ZhangDing15arxiv}. The linear fit of all the data points in which both $T_c$ and the maximum gap are known is plotted in red. For the systems in which only $T_c$ or gap is measured, the other missing quantity is calculated from the line of best fit. In the bottom of the figure are cartoons characterizing the three groups: with no charge transfer or substrate effect (left), with external charge transfer but no substrate effect (middle), and with substrate-induced charge transfer and e-ph coupling (right).
 }
\end{figure}
In Fig. 3 the maximum gap size vs $T_c$ for several iron chalcogenide superconductors are plotted. The relationship between gap and $T_c$ can be well fitted by a linear function. For the three systems in which only $T_c$ or the gap is measured, the other missing parameter is approximated using a fitted line. Three clear groupings emerge when displayed this way.

The bulk/multilayer films FeSe\cite{TanFeng13nmat,Hsu08pnas,NakayamaTakahashi15prb} and FeSe$_{(1-x)}$Te$_x$\cite{Noji10jps} are grouped at the bottom, with $T_c$ below 20 K. The 1-4 layer FeSe/SiC\cite{SongXue12prb} is included in this group as the interaction between SiC substrate and FeSe is believed to be weak, and this system shows similarly low $T_c$ and gap as bulk FeSe. Systems in this category are free of external doping and substrate effects.

The second group is distinguished by external electron doping, from either deposited K atoms or ionic liquid gating\cite{MiyataTakahashi15nmat,YeYan15arxiv,WenFeng16ncomm,TangXue15arxiv,LeiChen15arxiv,Shiogai15nphy}. All the systems in this category have a $T_c$ ranging from 31 K to 48 K, which is higher than all those belonging to the first group. ARPES measurements on K doped bulk FeSe and FeSe multilayer film on STO have shown a band structure similar to that of 1ML FeSe/STO$^{(001)}$, with only electron pocket(s) at M and no hole pocket at $\Gamma$. Several ARPES and STM works observed the coexistence of nematic order and superconducting phase, and a dome-like phase diagram in which $T_c$ can be tuned through electron doping\cite{WenFeng16ncomm,YeYan15arxiv,TangXue15arxiv}. The substrate effect is negligible for the materials in this category. This is supported by our results (Fig. 1f): no replica band can be observed in the K-doped 3ML FeSe/STO$^{(001)}$ film, although they share similar band structure.

The third group contains all 1ML FeSe films on titanate substrates. Despite substantial variations in strain, substrate dielectric constant and C4 symmetry, these 1ML FeSe films all have $T_c$ close to the temperature of liquid nitrogen and share similar electronic structure and doping levels \cite{Lee14nature,PengFeng14ncomm,ZhangDing15arxiv,ZhouXue15arxiv,DingXue16arxiv}. There is no significant difference in the Fermi surfaces between the monolayer films in this group and the electron doped systems in the second group. However, ARPES measurements on 1ML FeSe/STO$^{(001)}$ and 1ML FeSe/TiO$_2^{(100)}$ reveal the existence of replica bands, indicating strong interfacial e-ph interaction. Polaron bands on the two-dimensional electron gas (2DEG) have also been observed on TiO$_2$ anatase (001) and STO (110)\cite{Mocer13arxiv,Zhang16APS}, which suggests that a strong interfacial e-ph interaction is a common feature of the 1ML FeSe on oxide substrate systems, and thus is likely the cause of the $T_c$ enhancement.  

In conclusion, we used ARPES to compare the band structures of multilayer FeSe, K-doped multilayer FeSe, 1ML FeSe/STO$^{(001)}$ and 1ML FeSe/TiO$_2$ $^{(100)}$. Electron doping can enhance the $T_c$ of multilayer to between 30 K and 50 K, but only with interfacial e-ph coupling in 1ML FeSe can $T_c$ reach to about 70 K. 

The comparison between 1ML FeSe/STO$^{(001)}$ and 1ML FeSe/TiO$_2$ $^{(100)}$ indicates that among all substrate effects, the high electric constant, strain and lattice mismatch on superconductivity is limited, while interfacial e-ph coupling is the key reason for 1ML FeSe/STO$^{(001)}$ system to have a boosted $T_c$.

\begin{acknowledgments}
We would like to thank S. Johnston, B. Moritz, C.-J. Jia, J.-F. He, Y. He and S.-D. Chen for discussions. This work was supported by the U.S. Department of Energy, Office of Sciences, Basic Energy Sciences, Materials Sciences and Engineering Division.  ARPES measurements were performed at the Stanford Synchrotron Radiation Lightsource, a national user facility operated by Stanford University on behalf of the U.S. Department of Energy, Office of Basic Energy Sciences.  
\end{acknowledgments}

\end{document}